\documentstyle[epsfig,prb,aps,twocolumn]{revtex}

\begin{document}

\title{Switching the magnetic configuration of a spin valve by current induced domain wall motion}
\author{J. Grollier, D. Lacour, V. Cros, A. Hamzic\cite{Hamzic}, A. Vaur\`es,  A. Fert}
\address{Unit\'e Mixte de Physique CNRS-THALES, Domaine de Corbeville, 91404 Orsay Cedex, France
and Universit\'e Paris Sud, 91405 Orsay Cedex, France}
\author{D. Adam}
\address{THALES Research \& Technology France, Domaine de Corbeville, 91404 Orsay, France}
\author{G. Faini}
\address{Laboratoire de Photonique et de Nanostructures, LPN-CNRS, Route de Nozay, 91460 Marcoussis, France}

 \maketitle

\begin{abstract}
 We present experimental results on the displacement of a  domain wall by injection
 of a dc current through the wall. The samples are 1 $\mu$m wide long stripes of a CoO/Co/Cu/NiFe classical spin valve structure.
  The stripes have been patterned by electron beam lithography. A neck has been defined at 1/3 of the total length
  of the stripe and is a pinning center for the domain walls, as shown by the steps of the giant magnetoresistance curves at
  intermediate
  levels (1/3 or 2/3) between the resistances corresponding to the
parallel and antiparallel configurations. We show by electric
transport measurements that, once a wall is
  trapped, it can be moved by injecting a dc current higher than  a threshold current of the
  order of magnitude of 10$^7$ A/cm$^2$. We  discuss the different possible origins of this effect,
  i.e. local magnetic field created by the current and/or spin transfer from spin polarized current.
\end{abstract}

\vspace{.4in}

The conventional way to switch the magnetic configuration of a
spin electronic device is by generating a magnetic field with an
external current line. For submicronic devices, this has several
drawbacks in terms of energy consumption and risk of cross-talk.
A recently proposed alternative way rests on passing an
electrical current through the device to switch its magnetic
configuration, either by spin transfer from a spin polarized
current or by using the current-induced Oersted field. The
magnetization reversal of a small dot by spin transfer predicted
by Slonczewski\cite{Slonczewski} and Berger\cite{Berger pilier}
has now been confirmed by experiments on multilayered
pillars\cite{Katine1,Grollier} or nanowires\cite{Wegrowe}, and
magnetic switching by the current-induced Oersted field has also
been observed in other types of multilayered
pillars\cite{Bussmann,Katine2}.

In systems in which the magnetic configuration is defined by
domains separated by domain walls (DW), a possible mechanism of
magnetic switching is also the so-called current-induced domain
wall drag. There are several origins of the interaction between a
DW and an electrical current : the hydromagnetic drag force, which
arises from the Hall effect and is not significant for very thin
films \cite{hydromagnetic}, the current-induced field (Oersted
field) and the spin transfer by {\it s-d} interaction if the
current is spin-polarized. This last effect, predicted
theoretically by Berger\cite{Berger theorie}, has an origin
similar to the spin transfer mechanism referred to above. It
arises from the {\it s-d} exchange interaction between the spin
polarized electrons carrying the current and the local moments.
The {\it s-d} interaction exerts a torque on the spins of the
conduction electrons passing through a DW and rotates the
polarization direction of the current. Inversely, the spin
polarized current exerts a {\it s-d} exchange torque on the DW
magnetic configuration and thus can give rise to a motion of the
DW. The DW-drag by spin transfer can be significant for thin
enough DW in which the conduction electron spins cannot follow
completely the local magnetization direction. This condition can
be compared to the non-adiabatic criteria that has to be fulfilled
in order to observe DW magnetoresistance\cite{Viret}. Berger
{\it et al}.\cite{Berger experiences 1} have obtained some
experimental evidence of DW-drag  by injecting high dc current
pulses in thin films and observing DW position by Kerr
microscopy. The authors ascribe the DW-drag to {\it s-d}
exchange (spin transfer). In recent experiments on 100-160 nm
thick permalloy films, Gan {\it et al}.\cite{Gan} have also
observed DW displacement due to current pulses by imaging the DW
before and after the pulse using MFM. Their results suggest a
combination of spin transfer and hydromagnetic DW-drag. The key
points in these experiments are, first that the direction of the
DW displacement is reversed when the direction of  dc current
pulses is reversed, and second that the order of magnitude of the
current pulses needed to move the DW is always 10$^7$ A/cm$^2$.

\begin{figure}[bt]
 \centering \epsfig{file=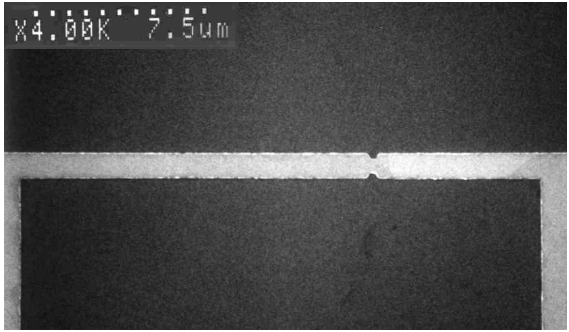,
width=8.5cm}
 \caption{SEM photography of the trilayer stripe and the neck. The width of the stripe is 1 $\mu$m and 0.5 $\mu$m in the constriction.}
 \label{fig1}
\end{figure}

The objective of the present work is to demonstrate that DW-drag
can be used to switch the magnetic configuration of a magnetic
device, a spin valve structure in this letter. We have used
sputtering and e-beam lithography to fabricate 1 $\mu$m wide and
20 $\mu$m long stripes of a CoO 30 $\AA$ /Co 70 $\AA$/Cu 100
$\AA$/NiFe 100 $\AA$ spin valve-type multilayered structure. A
constriction (0.5 $\mu$m wide neck) is also  patterned at one
third of the length, as shown in the SEM image of Fig. \ref{fig1}.
The depth of the notches is  0.25 $\mu$m and their shape is
approximately  triangular with a basis of about 0.3 $\mu$m. The
antiferromagnetic CoO layer is used to pin the magnetization of
the Co layer and to obtain this way well defined parallel and
antiparallel configurations by reversing the magnetization of the
soft permalloy layer (minor cycles). As the N\'eel temperature of
the antiferromagnet CoO is under 200 K, we have performed the
experiments at low temperature (3K). The stripe geometry with a
neck has proved to be efficient to trap a DW at the neck and to
detect its pinning and depinning directly by giant
magnetoresistance (GMR) measurements\cite{Ono}. The small width
of our stripes allows us to inject a high current density without
overheating and thus to avoid the use of current pulses in
contrast to ref.\cite{Berger experiences 1,Gan}. The resistance
is measured with a standard four contact dc technique, and a
magnetic field is applied along the long side of the stripe.

\begin{figure}[bt]
\centering \epsfig{file=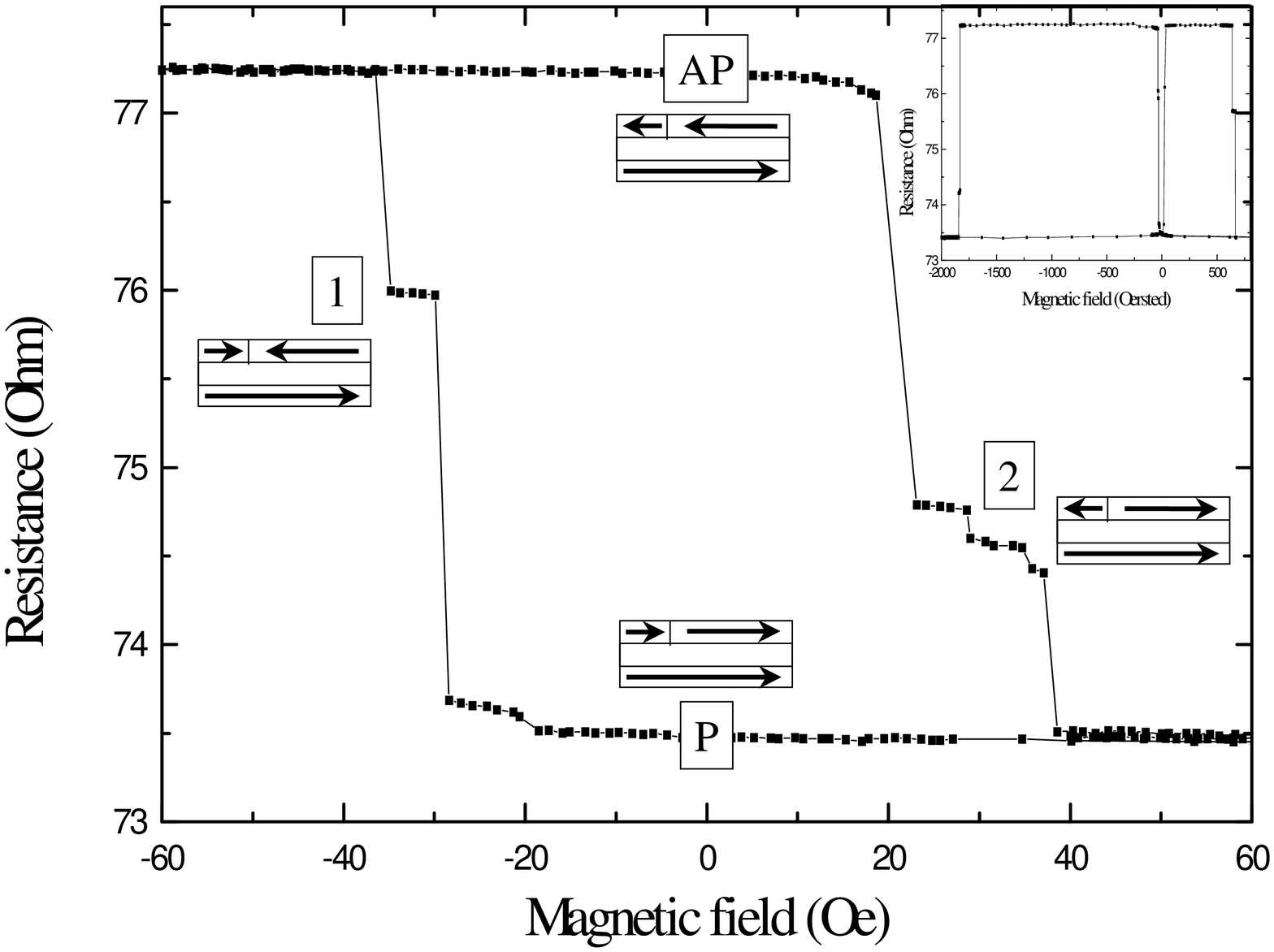, width=8.5cm}
 \caption{Magnetoresistance curve obtained at 3 K with the magnetic field applied along the stripe.
 The presented loop is a minor cycle showing the permalloy reversal,
 whereas the complete loop is plotted in the insert.}
 \label{fig2}
\end{figure}

 In Fig. \ref{fig2} we show an example of a GMR curve (a minor cycle, with the Co moment pinned in the positive field direction)
 for which the measuring current was 5 $\mu$A and the field resolution 1 Oe. The steps at intermediate levels (1/3 and 2/3)
 between the
   resistances of the parallel and antiparallel configurations are clearly seen. This is the proof that the DW is trapped at the  neck,
   as illustrated by the sketches on Fig. \ref{fig2}.

\begin{figure}[bt]
\centering \epsfig{file=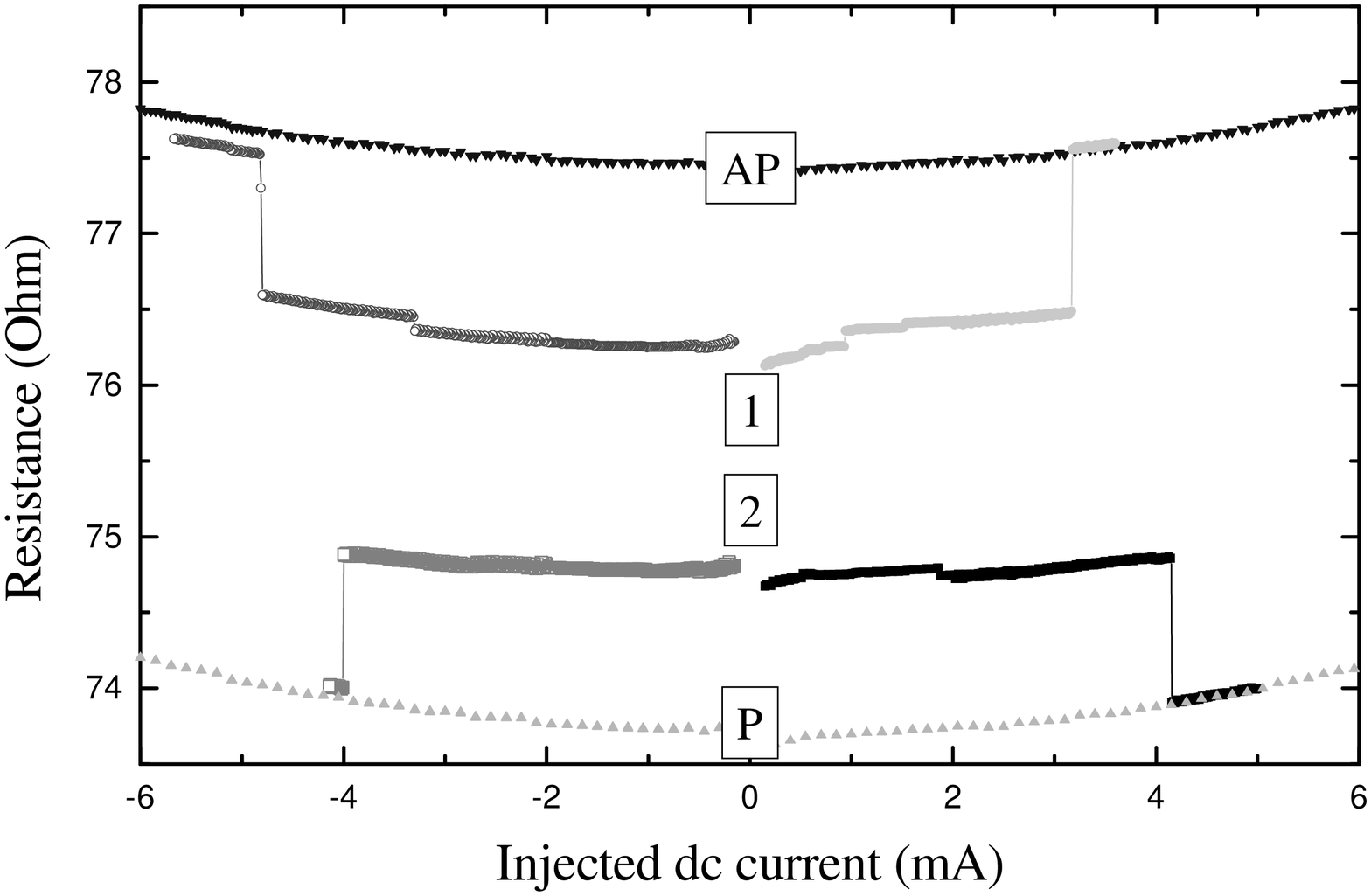, width=8.5cm}
 \caption{Resistance versus current curves. States 1 and 2 correspond to those indicated on the GMR curve of Fig.2.}
 \label{fig3}
\end{figure}

In order to study the effect of a dc current on the DW, the
following procedure was used. The field cycling is stopped at one
of the intermediate steps of the minor loop. In a first set of
experiments, we stop at state 1 corresponding to a - 27 Oe applied
magnetic field. Then, keeping the field constant, we increase or
decrease the current. The variation of the resistance as a
function of the current is shown in Fig. \ref{fig3}. The
resistance first remains practically at its initial value,
exhibiting only a slight reversible increase due to some heating
of the sample. By comparing this resistance increase to the
resistance versus temperature curve, we have estimated that the
maximum increase of temperature in our experiments does not
exceed 30 K, what, as we have checked\cite{temperature}, is
definitely insufficient to depin the DW. Then, when the current
reaches a threshold value (critical current) of about 4 mA, the
resistance jumps to the level corresponding to the AP
configuration, which is the more stable state in a negative
field. When the experiment is repeated starting from state 2 with
a + 27 Oe field, at the same threshold current, the resistance
jumps to the value of the stable P configuration.

We have added, for clarity, on Fig. \ref{fig3} the resistance
versus current curves obtained in the P and AP configurations and
the vertical jumps from an intermediate resistance level to the
levels of these two stable configurations are clearly observable.
These jumps are the signature of DW depinning and displacement
when the current exceeds a threshold value. We have also found
that, once the system is in the monodomain P or AP configuration,
it cannot be driven back to a pinned configuration (intermediate
level of resistance) by varying the current.

The absolute values of the depinning critical currents in
repeated experiments are scattered between 1.5 and 5 mA. It can
be pointed out that a  current of 4 mA corresponds to a density
of current crossing the neck of  2.6 10$^7$ A/cm$^2$, and 5
10$^6$ A/cm$^2$ if we consider only the current within  the NiFe
layer. This is in agreement with the order of magnitude given by
L. Berger {\it et al}. and L. Gan {\it et al}.\cite{Berger
experiences 1,Gan}. However a crucial point in our experiments is
that the effect is symmetric with respect to the sign of the
current, i.e. the DW is moved in the same direction for both
current directions (cf. Fig. \ref{fig3}). This is in contradiction
with the theoretical predictions\cite{Berger theorie} for DW-drag
by spin transfer (and would also be in contradiction with a
hydromagnetic mechanism, that, in any case, we do not consider for
our very thin layers). This is also in contrast with the recent
MFM observation of DW motion in permalloy films\cite{Gan}.

Another possible origin of DW-drag is the current-induced
(Oersted) field. The switching current of 4 mA should induce an
in-plane transverse field of a hundred Oe. This field is much
larger than the coercive field of the DW, but it has no component
along the stripe that could be added to the applied field and
directly contribute to the depinning. The longitudinal component
of the Oersted field is in average zero in the neck, but can
reach local values up to a few tenth of Oe, due to the neck
geometry. This leads us to consider a possible twist and
destabilization of the DW related to the inhomogeneity of the
current-induced field. It should also be emphasized that a similar
DW twist and depinning induced by the inhomogeneity of the spin
transfer torque cannot be ruled out. In other words, for the
specific geometry of the constriction, domain drag by spin
transfer could also present different features than for DW in
standard films.

In conclusion, we have shown that a dc current can switch the
magnetic configuration of a spin valve structure by displacing a
domain wall pinned by a constriction. The origin of the effect is
not yet clearly established: we are not able to explain our
results neither by the spin transfer model worked out for standard
DW, nor by the effect of the field generated by the current.
Experiments with smaller constrictions should be useful to
discriminate the two mechanisms. On the other hand, from a purely
technological point of view, our finding of current-induced
switching in a spin valve device indicates a promising way to
control the spin electronic devices. Switching back and forth the
configuration of a device by moving a domain wall between two
constrictions in a nanosecond time scale should be the next step
in this direction.

This work was supported by the EU through the RTN "Computational
Magnoelectronics" (HPRN-CT-2000-00143) and the Minist\`ere de la
Recherche et de la Technologie through the MRT "Magmem II"
(01V0030) and the ACI contract "BASIC" (27-01).


\begin{references}

\bibitem[a]{Hamzic} on leave from the Department of Physics, Faculty of
Science, HR-1000 Zagreb, Croatia.

\bibitem{Slonczewski}
J. Slonczewski, J. Magn. Magn. Mat. {\bf 159},  1  (1996).

\bibitem{Berger pilier}
L. Berger, Phys. Rev. B {\bf 54},  9353  (1996).

\bibitem{Katine1}
J.~A. Katine, F. J. Albert, R. A. Buhrman, E. B. Myers, D. C.
Ralph, Phys. Rev. Lett. {\bf 84}, 3149 (2000); F.~J. Albert, J.~A.
Katine, R.~A. Buhrman, and D.~C. Ralph, Appl. Phys. Lett.
  {\bf 77},  3809  (2000).

\bibitem{Grollier}
J. Grollier, V. Cros, A. Hamzic, J.M. George, H. Jaffr\`es, A.
Fert, G. Faini, J. Ben Youssef, H. Legall, Appl. Phys. Lett. {\bf
78}, 3663 (2001).

\bibitem{Wegrowe}
J.-E. Wegrowe, A. F\'abi\'an, Ph. Guittienne, X. Hoffer, D.
Kelly, J.-Ph. Ansermet, E. Olive, Appl. Phys. Lett. {\bf 80},
   3775  (2002).

\bibitem{Bussmann}
K. Bussmann, G.~A. Prinz, S.~F. Cheng, and D. Wang, Appl. Phys.
Lett. {\bf 75},
   2476  (1999).

\bibitem{Katine2}
J.~A. Katine, F.~J. Albert, and R.~A. Buhrman, Appl. Phys. Lett.
{\bf 76},  354
   (2000).

\bibitem{hydromagnetic}
W. J. Carr, J. Appl. Phys {\bf 45}, 394 (1974). L. Berger, J.
Phys. Chem. Solids {\bf 35}, 947 (1974).

\bibitem{Berger theorie}
L. Berger, J. Appl. Phys.  {\bf 55},  1954  (1984); L. Berger, J.
Appl. Phys.  {\bf 71},  2721  (1992).

\bibitem{Viret}
M. Viret, D. Vignoles, D. Cole, J. M. D. Coey, W. Allen, D. S.
Danile, J. F. Gregg, Phys. Rev. B {\bf 53}, 8464  (1996).

\bibitem{Berger experiences 1}
P.P. Freitas, L. Berger, J. Appl. Phys.  {\bf 57},  1266  (1985);
C.-Y. Hung, L. Berger, J. Appl. Phys.  {\bf 63},  4276  (1988);
C.-Y. Hung, L. Berger, C. Y. Shih, J. Appl. Phys.  {\bf 67},  5941
(1990); E. Salhi, L. Berger, J. Appl. Phys.  {\bf 73},  6405
(1993); E. Salhi, L. Berger, J. Appl. Phys.  {\bf 76},  4787
(1994).

\bibitem{Gan}
L. Gan, S. H. Chung, K. H. Aschenbach, M. Dreyer, R. D. Gomez,
IEEE Trans. Mag. {\bf 36}, 3047 (2000).

\bibitem{Ono}
T. Ono, H. Miyajima, K. Shigeto, T. Shinjo, Appl. Phys. Lett. {\bf
72}, 1116 (1998).


\bibitem{temperature}
The DW was pinned at state 1 at 3K and with a - 27 Oe magnetic
field. Then, at constant field, the temperature was increased up
to 150 K then down again to 3 K. The resistance at the end of
this temperature cycle was found to be unchanged from that
obtained at state 1 at the beginning of the experiment.







\end{references}
\end {document}